\documentclass[12pt]{article}

\usepackage{amsmath,amsthm,amssymb,a4wide}

\usepackage{mybbold}    

\newcommand{\ud}{\text{d}}
\newcommand{\ui}{\text{i}}
\newcommand{\ue}{\text{e}}
\newcommand{\SU}{\mathrm{SU}}

\newcommand{\al}{\alpha}
\newcommand{\be}{\beta}
\newcommand{\de}{\delta}

\newcommand{\ga}{\gamma}

\newcommand{\om}{\omega}
\newcommand{\Om}{\Omega}
\newcommand{\si}{\sigma}

\newcommand{\ve}{\varepsilon}
\newcommand{\vp}{\varphi}

\newcommand{\cH}{{\mathcal H}}

\newcommand{\cS}{{\mathcal S}}

\newcommand{\vecA}{\boldsymbol{A}} 
\newcommand{\vecB}{\boldsymbol{B}}

\newcommand{\vecE}{\boldsymbol{E}}

\newcommand{\vecx}{\boldsymbol{x}}
\newcommand{\vecX}{\boldsymbol{X}}
\newcommand{\vecy}{\boldsymbol{y}}
\newcommand{\vecz}{\boldsymbol{z}}
\newcommand{\vecp}{\boldsymbol{p}}
\newcommand{\vecP}{\boldsymbol{P}}

\newcommand{\vecal}{\boldsymbol{\alpha}} 
\newcommand{\vecsig}{\boldsymbol{\sigma}}

\newcommand{\vecxi}{\boldsymbol{\xi}}

\newcommand{\rz}{{\mathbb R}}
\newcommand{\nz}{{\mathbb N}}
\newcommand{\kz}{{\mathbb C}}
\newcommand{\mz}{{\mathbb M}}

\DeclareMathOperator{\tr}{Tr}
\DeclareMathOperator{\mtr}{tr}
\DeclareMathOperator{\vol}{vol}        
\DeclareMathOperator{\supp}{supp}     

\newcommand{\eins}{\mathmybb{1}}

\numberwithin{equation}{section}

\begin{document}

\thispagestyle{empty}

\noindent
ULM-TP/00-5\\
September 2000\\

\vspace*{1cm}

\begin{center}

{\LARGE\bf Semiclassical expectation values for \\
\vspace*{3mm}
relativistic particles with spin 1/2} \\
\vspace*{3cm}
{\large Jens Bolte}%
\footnote{E-mail address: {\tt bol@physik.uni-ulm.de}} 

\vspace*{1cm}

Abteilung Theoretische Physik\\
Universit\"at Ulm, Albert-Einstein-Allee 11\\
D-89069 Ulm, Germany 
\end{center}

\vfill

\begin{abstract}
For relativistic particles with spin 1/2, which are described by the Dirac
equation, a semiclassical trace formula is introduced that incorporates 
expectation values of observables in eigenstates of the Dirac-Hamiltonian.
Furthermore, the semiclassical limit of an average of expectation values
is expressed in terms of a classical average of the corresponding classical
observable.  
\end{abstract}

\newpage

\section{Introduction}
\label{1sec}
The three decades that passed since the completion of Gutzwiller's 
pioneering work on the trace formula \cite{Gut71} (see also \cite{Gut90})
have witnessed a flourishing development of semiclassical methods in 
connection with trace formulae. Gutzwiller himself, e.g., applied his 
semiclassical representation of the spectral density in terms of a sum
over classical periodic orbits to the quantisation of the anisotropic 
Kepler problem \cite{Gut71,Gut73}. Later, in the context of the then 
developing field of quantum chaos, it was realised that the Gutzwiller 
trace formula can quite generally be employed for a semiclassical
quantisation of classically chaotic systems, where no Bohr-Sommerfeld
(or Einstein-Brillouin-Keller) type quantisation is available. A number
of semiclassical quantisation schemes were invented and successfully 
passed various tests, see \cite{Gut90,Chaos2,AurMatSieSte92} for collections 
of examples. On the other hand, Hannay and Ozorio de Almeida \cite{HanOzo84} 
and Berry \cite{Ber85} realised that a semiclassical analysis of spectral 
correlations can be based on trace formulae. These techniques, which
have meanwhile been refined in several aspects, still provide the strongest
analytic support for the conjectured universality in energy level statistics
\cite{BohGiaSch84}. 

Shortly after Gutzwiller established his trace formula, and apparently
independently thereof, mathematicians began to prove various versions of
semiclassical trace formulae \cite{CdV73,DuiGui75,Mei92,PauUri95,ComRalRob99}. 
Moreover, generalisations of the trace formula as a semiclassical 
representation of the spectral density were developed, e.g., to include also 
expectation values of observables \cite{Wil88,PauUri95}. Thus, by now the 
Gutzwiller trace formula is a well developed and well established tool in 
semiclassical quantum mechanics.

So far, trace formulae have been derived for quantum systems with only
translational degrees of freedom. These possess a well defined classical
limit and thus semiclassical methods can be applied in a straight forward
manner. From a physical point of view, however, spin degrees of freedom
are of particular importance. Quantum mechanically a spin 1/2 of, say,
an electron is described in a two- (i.e.\ finite-) dimensional Hilbert
space, lacking a direct classical counterpart, so that a semiclassical 
description of quantum systems with coupled translational and spin degrees 
of freedom is less obvious than in the previous cases. However, the 
semiclassical formalism that was primarily developed in the mathematical 
community (see, e.g., \cite{DuiGui75,Mei92,PauUri95}) can also be applied 
to derive a trace formula for Dirac- and Pauli-Hamiltonians 
\cite{BolKep98,BolKep99}. For these, alike in Gutzwiller's original trace 
formula, the spectral density can be expressed in terms of a sum over 
classical periodic orbits (of the translational motion). Spin enters through 
weight factors that depend on a phase derived from a `classical' spin 
precessing along the classical periodic orbits.

It is the goal of this paper to develop a semiclassical theory for
expectation values of observables in the case of relativistic particles with 
spin 1/2. To this end we first extend the trace formula 
\cite{BolKep98,BolKep99} for Dirac-Hamiltonians to also include expectation 
values and then discuss averages of expectation values in eigenstates of the
Hamiltonian that correspond to a certain spectral interval. In this way
average expectation values and fluctuations thereabout are expressed in
terms of classical quantities. The article is organised as follows:
In section~2 we recall spectral properties of Dirac-Hamiltonians and 
represent them as Weyl operators. Bounded observables are then introduced 
in section~3. Subsequently, in section~4 the semiclassical description of 
the time evolution operator in terms of a Van Vleck-Gutzwiller propagator 
for Dirac-Hamiltonians that was developed in \cite{BolKep98,BolKep99} is 
discussed. Section~5 is then devoted to the derivation of the trace 
formula for the spectral density weighted by expectation values. Finally,
in section~6 we calculate averages of expectation values semiclassically.
\section{The Dirac-Hamiltonian}
\label{2sec}
In relativistic quantum mechanics a point particle of mass $m$ and
charge $e$ that is exposed to external electromagnetic fields $\vecE$
and $\vecB$ is described by the Dirac equation. Here we fix an inertial
frame and work with fixed coordinates $(t,\vecx)$ for Minkowski
space-time. We moreover assume that in the given frame the fields 
are static so that the Dirac equation reads
\begin{equation}
\label{Diraceq}
\ui\hbar\frac{\partial\psi}{\partial t}(t,\vecx)=\hat H\psi(t,\vecx)\ ,
\end{equation}
with the Dirac-Hamiltonian
\begin{equation}
\label{DiracHam}
\hat H = c\vecal\cdot\Bigl(\frac{\hbar}{\ui}\nabla -\frac{e}{c}\vecA(\vecx)
\Bigr) + mc^2\be + e\vp(\vecx) \ ,
\end{equation}
where $\vp,\vecA$ are electromagnetic potentials so that $\vecE(\vecx)
=-\nabla\vp(\vecx)$ and $\vecB(\vecx)=\nabla\times\vecA(\vecx)$. The Dirac
algebra $\al_\mu\al_\nu +\al_\nu\al_\mu =2\de_{\mu\nu}$, $\mu,\nu=0,\dots,3$, 
is realised by the hermitian $4\times 4$ matrices
\begin{equation}
\label{Diracalg}
\vecal = \begin{pmatrix} 0&\vecsig \\ \vecsig&0 \end{pmatrix} \qquad
\text{and}\qquad\al_0 = \be = \begin{pmatrix} \eins_2&0 \\ 0&-\eins_2
\end{pmatrix} \ ,
\end{equation}
where $\si_k$, $k=1,2,3$, are the Pauli matrices and $\eins_n$ denotes the
$n\times n$ unit matrix. For more details see \cite{Tha92}.

If the potentials $\vp,A_k$ are smooth functions, or at most have Coulomb
singularities at some points $\vecx_j$ such that
\begin{equation}
\label{esssacond}
|e\vp(\vecx)| ,\ |eA_k(\vecx)| \leq \frac{a}{|\vecx-\vecx_j|}+b
\end{equation}
holds for some constants $a<\frac{\hbar c}{2}$, $b>0$ and for all $\vecx\in
\rz^3\backslash\{\vecx_j\}$, the Dirac-Hamiltonian (\ref{DiracHam})
is essentially self-adjoint on the domain $C^\infty_0(\rz^3\backslash
\{\vecx_j\})\otimes\kz^4$ in the Hilbert space $\cH=L^2(\rz^3)\otimes\kz^4$
of square-integrable Dirac spinors (see, e.g., \cite{Tha92}). The quantum 
Hamiltonian is then given by the (unique) self-adjoint extension which we 
also denote by $\hat H$. As an example consider an electron in the Coulomb 
potential of a nucleus with charge $Ze$, such that 
$e\vp(\vecx)=-\frac{Ze^2}{|\vecx|}$. Since the fine structure constant 
$\al=\frac{e^2}{\hbar c}$ is approximately given by $\frac{1}{137}$, the 
bound (\ref{esssacond}) implies essential self-adjointness for $Z\leq 68$. 
In the case $\vecA=0$ and $\vp$ spherically symmetric, however, the bound 
on $a$ can be improved by a factor of $\sqrt{3}$, leading to $Z\leq 118$. 
For larger $Z$ one has to choose special self-adjoint extensions, see
\cite{Tha92} for further details.

Dirac-Hamiltonians (\ref{DiracHam}) possess spectra that are not bounded 
from below. In fact, if the electromagnetic fields vanish as 
$|\vecx|\to\infty$, the essential spectrum of $\hat H$ is given by
\begin{equation}
\label{essspec}
\si_{ess}(\hat H) = (-\infty,-mc^2] \cup [+mc^2,+\infty) \ .
\end{equation}
(This is also true under appropriate weaker conditions, see \cite{Tha92}.) 
In what follows we will always consider situations where $\hat H$ is 
(essentially) self-adjoint and where (\ref{essspec}) holds. Since below we 
are only interested in properties of the discrete spectrum of 
Dirac-Hamiltonians, we will therefore concentrate on the eigenvalues
$E_n$ of $\hat H$ with $|E_n|<mc^2$. More precisely, we choose some energy 
$E$ in the gap of $\si_{ess}(\hat H)$ and introduce  an interval
\begin{equation}
\label{Idef}
I(E;\hbar):=[E-\hbar\om,E+\hbar\om]\ ,\qquad \om >0 \quad\text{fixed}\ ,
\end{equation}
that has no overlap with the essential spectrum of $\hat H$. This only 
requires $\hbar$ to be sufficiently small. We are then interested in the 
eigenvalues $E_n\in I(E;\hbar)$ and the expectation values 
$\langle\psi_n,\hat B\psi_n\rangle$ of observables $\hat B$ in eigenstates 
$\psi_n$ related to these eigenvalues.

For the purpose of semiclassical asymptotics it is advantageous to view
the Hamiltonian, as well as all further observables, as Weyl operators.
For simplicity we now assume that the potentials are smooth so that the
Weyl representation
\begin{equation}
\label{HWeyl}
(\hat H\psi)(\vecx) = \frac{1}{(2\pi\hbar)^3}\int_{\rz^3}\int_{\rz^3}
H\Bigl(\vecp,\frac{\vecx+\vecy}{2}\Bigr)\,\ue^{\frac{\ui}{\hbar}\vecp
\cdot(\vecx-\vecy)}\,\psi(\vecy)\ \ud p\,\ud y \ 
\end{equation}
of $\hat H$ holds for $\psi\in C_0^\infty(\rz^3)\otimes\kz^4$. Here
\begin{equation}
\label{W-symb}
\begin{split}
H(\vecp,\vecx)&:=c\vecal\cdot\Bigl(\vecp-\frac{e}{c}\vecA(\vecx)\Bigr)+mc^2\be
                 +e\vp(\vecx) \\
              &=\begin{pmatrix} (e\vp(\vecx)+mc^2)\eins_2 & 
                (c\vecp-e\vecA(\vecx))\cdot\vecsig \\ 
                (c\vecp-e\vecA(\vecx))\cdot\vecsig & 
                (e\vp(\vecx)-mc^2)\eins_2
                \end{pmatrix}
\end{split} 
\end{equation}
is the Weyl symbol of $\hat H$, which for all $(\vecp,\vecx)\in\rz^3\times
\rz^3$ is a hermitian $4\times 4$ matrix with the two twofold degenerate
eigenvalues
\begin{equation}
\label{symbev}
H^\pm(\vecp,\vecx) = e\vp(\vecx)\pm\sqrt{(c\vecp-e\vecA(\vecx))^2+m^2c^4}\ ,
\end{equation}
such that $H^+ -H^-\geq 2mc^2 >0$. These eigenvalues are the classical
Hamiltonians for relativistic particles (without spin) with positive and
negative kinetic energies, respectively, that are exposed to the 
electromagnetic fields generated by the potentials $\vp$ and $\vecA$. The 
presence of these two eigenvalues is a remnant of the fact that the Dirac 
equation (\ref{Diraceq}) describes both particles and anti-particles, and
the dimension of the corresponding eigenspaces follows from the spin 1/2 
of the particles and anti-particles, respectively.

One can now introduce an orthonormal basis of $\kz^4$ consisting of the 
eigenvectors $e_1(\vecp,\vecx),e_2(\vecp,\vecx)$ of $H(\vecp,\vecx)$ with 
eigenvalue $H^+(\vecp,\vecx)$ and the eigenvectors
$f_1(\vecp,\vecx),f_2(\vecp,\vecx)$ with eigenvalue $H^-(\vecp,\vecx)$.
This basis allows to construct the $4\times 2$ matrices $V_+(\vecp,\vecx)$ 
and $V_-(\vecp,\vecx)$ with $e_1(\vecp,\vecx),e_2(\vecp,\vecx)$ and 
$f_1(\vecp,\vecx),f_2(\vecp,\vecx)$, respectively, as columns 
\cite{BolKep98,BolKep99},
\begin{equation}
\label{VWdef}
\begin{split}
V_+(\vecp,\vecx)&=\frac{1}{\sqrt{2\ve(\vecp,\vecx)(\ve(\vecp,\vecx)+mc^2)}}
                  \begin{pmatrix}
                  (\ve(\vecp,\vecx)+mc^2)\eins_2 \\ (c\vecp-e\vecA(\vecx))
                  \cdot\vecsig
                  \end{pmatrix}\ , \\
V_-(\vecp,\vecx)&=\frac{1}{\sqrt{2\ve(\vecp,\vecx)(\ve(\vecp,\vecx)+mc^2)}}
                  \begin{pmatrix}
                  (c\vecp-e\vecA(\vecx))\cdot\vecsig \\ 
                  -(\ve(\vecp,\vecx)+mc^2)\eins_2 
                  \end{pmatrix} \ ,
\end{split}
\end{equation}
where
\begin{equation}
\label{epsdef}
\ve(\vecp,\vecx):=\sqrt{(c\vecp-e\vecA(\vecx))^2 +m^2c^4} \ .
\end{equation}
The columns of $V_\pm$ being orthonormal implies that these matrices
represent isometries from $\kz^2$ into $\kz^4$ so that $V_\pm^\dagger V_\pm
=\eins_2$. Moreover, the projectors $P_0^\pm (\vecp,\vecx)$ onto the 
eigenspaces corresponding to the eigenvalues $H^\pm(\vecp,\vecx)$ are given 
by $P_0^\pm=V_\pm V_\pm^\dagger$. 
\section{Observables}
\label{3sec}
The kind of observables we are going to consider are given by matrix
valued Weyl operators as in (\ref{HWeyl}), i.e. 
\begin{equation}
\label{BWeyl}
(\hat B\psi)(\vecx) = \frac{1}{(2\pi\hbar)^3}\int_{\rz^3}\int_{\rz^3}
B\Bigl(\vecp,\frac{\vecx+\vecy}{2};\hbar\Bigr)\,\ue^{\frac{\ui}{\hbar}\vecp
\cdot(\vecx-\vecy)}\,\psi(\vecy)\ \ud p\,\ud y \ ,
\end{equation}
that yield bounded and self-adjoint operators on $\cH$. Notice that here
the symbol $B(\vecp,\vecx;\hbar)$ is allowed to depend on $\hbar$. In
order to ensure that the operator $\hat B$ defined by (\ref{BWeyl}) is
bounded one can impose the following condition on its symbol: it shall 
be smooth in the variables $(\vecp,\vecx)$ and, moreover, possess an
asymptotic expansion
\begin{equation}
\label{asymptot}
B(\vecp,\vecx;\hbar)\sim\sum_{j\geq 0}\hbar^j\,B_j(\vecp,\vecx) 
\end{equation}
in the sense that for all $N\geq 0$ and all multi-indices $\al,\be\in\nz_0^3$
\begin{equation}
\label{asympdef}
\Bigl\|\partial^\al_p\partial^\be_x \Bigl( B(\vecp,\vecx;\hbar)-\sum_{j=0}^N
\hbar^j\,B_j(\vecp,\vecx) \Bigr) \Bigr\|_{4\times 4} \leq C_{\al\be}\,
\hbar^{N+1} \ .
\end{equation}
In particular, $B$ and all of its derivatives shall be bounded.
Here $\|\dots\|_{4\times 4}$ denotes an arbitrary matrix norm and the 
short-hand $\partial^\al_x=\frac{\partial^{\al_1+\al_2+\al_3}}
{\partial x_1^{\al_1}\partial x_2^{\al_2}\partial x_3^{\al_3}}$ has been 
used. The leading term $B_0(\vecp,\vecx)$ in the semiclassical asymptotics
(\ref{asymptot}) is called {\it principal symbol} and will be viewed
as the classical observable corresponding to the quantum observable $\hat B$.
For scalar symbols a proof for the fact that (\ref{asympdef}) implies the 
boundedness of $\hat B$ can, e.g., be found in \cite{DimSjo99}; this
immediately carries over to the present situation, see also \cite{BolGla00}.
The operator $\hat B$ defined by (\ref{BWeyl}) for $\psi\in C_0^\infty(\rz^3)
\otimes\kz^4$ hence can now be extended to all of $\cH$ and is self-adjoint
as soon as its symbol $B(\vecp,\vecx;\hbar)$ is hermitian. In the 
following we will always assume this to be the case.

Two particular classical observables are the projectors $P_0^\pm(\vecp,\vecx)$
onto the eigenspaces of the symbol (\ref{W-symb}) of the Dirac-Hamiltonian.
If we assume that the potentials and all of their derivatives are bounded,
which poses no severe restriction since we want the electromagnetic
fields to vanish as $|\vecx|\to\infty$, an inspection of (\ref{VWdef})
reveals that $P_0^\pm(\vecp,\vecx)$ together with all derivatives are
also bounded. A Weyl quantisation of $P_0^\pm(\vecp,\vecx)$ hence yields 
bounded operators $\hat P_0^\pm$. These, however, are not yet projection 
operators on $\cH$, but a semiclassical construction, see 
\cite{EmmWei96,BruNou99}, allows to introduce bounded and self-adjoint 
operators $\hat P^\pm$ with principal symbols $P^\pm_0(\vecp,\vecx)$ that 
obey $\hat P^+ +\hat P^- =\eins_\cH$. Moreover, up to terms of order 
$\hbar^\infty$ these operators fulfill $(\hat P^\pm)^2 = \hat P^\pm$, 
$\hat P^+\hat P^- =0$ and commute with $\hat H$. With respect to these 
semiclassical projectors one can now split an observable $\hat B$ of the 
type introduced above according to
\begin{equation}
\label{Bsplit}
\hat B = \hat B_d +\hat B_{nd}\ ,
\end{equation}
where
\begin{equation}
\label{Bsplitdef}
\hat B_d := \hat P^+\hat B\hat P^+ + \hat P^-\hat B\hat P^-\qquad\text{and}
\qquad \hat B_{nd} := \hat P^+\hat B\hat P^- + \hat P^-\hat B\hat P^+
\end{equation}
denote the diagonal and the non-diagonal part, respectively. According
to the product rule for Weyl operators (see, e.g., \cite{Rob87,DimSjo99})
$\hat B_d$ and $\hat B_{nd}$ are again bounded and self-adjoint Weyl 
operators with principal symbols
\begin{equation}
\label{Bsplitpsymb}
\begin{split}
B_{d,0}(\vecp,\vecx) &= P_0^+(\vecp,\vecx)\,B_0(\vecp,\vecx)\,
                        P_0^+(\vecp,\vecx)+P_0^-(\vecp,\vecx)\,
                        B_0(\vecp,\vecx)\,P_0^-(\vecp,\vecx) \ ,\\
B_{nd,0}(\vecp,\vecx)&= P_0^+(\vecp,\vecx)\,B_0(\vecp,\vecx)\,
                        P_0^-(\vecp,\vecx)+P_0^-(\vecp,\vecx)\,
                        B_0(\vecp,\vecx)\,P_0^+(\vecp,\vecx) \ .
\end{split}
\end{equation}
\section{Semiclassical time evolution}
\label{4sec}
In this section we mainly review the derivation of the Van Vleck-Gutzwiller 
propagator for Dirac-Hamiltonians that was developed in 
\cite{BolKep98,BolKep99}. For a detailed exposition see \cite{BolKep99}.
The aim here is to obtain the semiclassically leading asymptotics of the 
Schwartz kernel $K(\vecx,\vecy,t)$ for the unitary time evolution operator 
$\hat U(t)=\ue^{-\frac{\ui}{\hbar}t\hat H}$. Given an initial state 
$\psi_0\in\cH$, the solution $\psi(t)=\hat U(t)\psi_0$ of the Dirac equation 
(\ref{Diraceq}) with initial condition $\psi(0)=\psi_0$ reads in coordinate 
representation
\begin{equation}
\label{kernel}
\psi(t,\vecx) = \int_{\rz^3}K(\vecx,\vecy,t)\,\psi_0(\vecy)\ \ud y \ ,
\end{equation}
so that the initial condition for the kernel is $K(\vecx,\vecy,0)=
\de(\vecx-\vecy)$. Up to terms of order $\hbar^\infty$ this kernel can
be approximated by a semiclassical Fourier integral operator, i.e.
\begin{equation}
\label{scFIO}
\begin{split}
K(\vecx,\vecy,t)=\frac{1}{(2\pi\hbar)^3}\int_{\rz^3}
&\left[ a_\hbar^+(\vecx,\vecy,t,\vecxi)\,\ue^{\frac{\ui}{\hbar}\phi^+
 (\vecx,\vecy,t,\vecxi)}\right. \\
&\left. +a_\hbar^- (\vecx,\vecy,t,\vecxi)\,\ue^{\frac{\ui}{\hbar}\phi^-
 (\vecx,\vecy,t,\vecxi)} \right]\ \ud\xi + O(\hbar^\infty) \ .
\end{split}
\end{equation}
The two additive contributions under the integral refer to the two 
eigenvalues $H^\pm$ of the symbol $H$ of the Dirac-Hamiltonian, which each 
serve as a classical Hamiltonian. The amplitude factors $a_\hbar^\pm$
take values in the $4\times 4$ matrices and are assumed to allow for the
asymptotic expansions
\begin{equation}
\label{ampexpand}
a^\pm_\hbar (\vecx,\vecy,t,\vecxi) = \sum_{k\geq 0} \Bigl(\frac{\hbar}{\ui}
\Bigr)^k\,a^\pm_k (\vecx,\vecy,t,\vecxi) \ ,
\end{equation}
with initial conditions
\begin{equation}
\label{ampincond}
a^\pm_k (\vecx,\vecy,0,\vecxi) = 
  \begin{cases}
  P_0^\pm(\vecxi,\vecx) & \text{if $k=0$}\ , \\
  0 & \text{if $k\geq 1$}\ .
  \end{cases}
\end{equation}
Together with the initial condition $\phi^\pm|_{t=0}=\vecxi\cdot(\vecx-\vecy)$
for the phase function this ensures the desired behaviour of the 
representation (\ref{scFIO}) for $t\to 0$.

The strategy to obtain a semiclassical approximation for $K(\vecx,\vecy,t)$
now is the same as in the WKB method: one inserts the ansatz (\ref{scFIO}) 
with the expansion (\ref{ampexpand}) into the Dirac equation, groups terms 
together with like powers of $\hbar$, and equates these terms to zero. This 
yields a hierarchy of equations for the phase functions $\phi^\pm$ and the 
coefficients $a_k^\pm$ that have to be solved order by order in $k$. The
lowest order ($k=0$) equation, together with the definition 
$S^\pm(\vecxi,\vecx,t):=\phi^\pm(\vecx,\vecy,t,\vecxi)+\vecxi\cdot\vecy$, 
relates the phase functions $\phi^\pm$ to the solutions $S^\pm$ of the 
associated Hamilton-Jacobi equations
\begin{equation}
\label{HJeq}
H^\pm\bigl(\nabla_x S^\pm(\vecxi,\vecx,t),\vecx\bigr) + \frac{\partial S^\pm}
{\partial t}(\vecxi,\vecx,t) = 0 \qquad\text{with}\quad S^\pm(\vecxi,\vecx,0)
=\vecxi\cdot\vecx \ .
\end{equation}
Hence $S^\pm$ generate canonical transformations $(\vecp^\pm,\vecx)\mapsto
(\vecxi,\vecz^\pm)$ with $\vecp^\pm =\nabla_x S^\pm(\vecxi,\vecx,t)$ and
$\vecz^\pm =\nabla_\xi S^\pm(\vecxi,\vecx,t)$ such that $(\vecxi,\vecz^\pm)$ 
and $(\vecp^\pm,\vecx)$ are starting and end points, respectively, of 
solutions $(\vecP^\pm(t'),\vecX^\pm(t'))$, $0\leq t'\leq t$, of the classical
equations of motion generated by $H^\pm$. 

The leading semiclassical asymptotics for the amplitudes then follows from
both the equations in leading and next-to-lowest order. For sufficiently small 
$t$ the result reads 
\begin{equation}
\label{a0sol}
\begin{split}
a_0^\pm (\vecx,\vecy,t,\vecxi)= 
&\sqrt{\det\Bigl(\frac{\partial^2 S^\pm}{\partial\xi\partial x} 
 (\vecxi,\vecx,t)\Bigr)} \\
&V_\pm\bigl(\nabla_x S^\pm(\vecxi,\vecx,t),\vecx\bigl)\,\tilde d_\pm
 (\nabla_x S^\pm(\vecxi,\vecx,t),\vecx)\,V_\pm^\dagger (\vecxi,\vecy) \ . 
\end{split}
\end{equation}
The restriction to small $t$ stems from the non-uniqueness of the solutions 
$S^\pm$ of the Hamilton-Jacobi equations (\ref{HJeq}) for times beyond which 
caustics appear. One is, however, faced with the same problem already in 
the context of non-relativistic, spinless particles where the respective 
solution merely consists of the first factor on the right-hand side of 
(\ref{a0sol}). For this case Gutzwiller showed \cite{Gut67} that one can 
pass beyond caustics with a correct choice of the phase in the first term of
(\ref{a0sol}) when sign changes occur under the square-root. The appropriate
phase factor that Gutzwiller introduced essentially consists of the Morse 
index of the related classical trajectory. A mathematically rigorous 
justification of this construction can, e.g., be found in 
\cite{DuiGui75,Mei92}. Obviously, exactly the same procedure can be applied 
to extend (\ref{a0sol}) to arbitrary, however finite, times.

The factor $\tilde d_\pm$ appearing in (\ref{a0sol}) is a solution of the 
spin transport equation
\begin{equation}
\label{spintranseq}
\Bigl( \frac{\ud}{\ud t}+\ui M_\pm(\vecP^\pm (t),\vecX^\pm (t)) \Bigr)\, 
\tilde d_\pm(\vecP^\pm(t),\vecX^\pm(t)) = 0 \ ,
\end{equation}
with initial condition $\tilde d_\pm(\vecP^\pm(0),\vecX^\pm(0))=\eins_2$, 
in which the time derivative is to be taken along the solutions
$(\vecP^\pm (t),\vecX^\pm (t))$ of the classical equations of motion that
are generated by the solutions $S^\pm$ of the Hamilton-Jacobi equations
(\ref{HJeq}). The $2\times 2$ matrices
\begin{equation}
\label{Mpmdef}
M_\pm(\vecp,\vecx) := \mp\frac{ec}{2\ve(\vecp,\vecx)}\left[\vecB(\vecx)
\pm\frac{c}{\ve(\vecp,\vecx)+mc^2}\Bigl(\vecE(\vecx)\times\bigl(\vecp -
\frac{e}{c}\vecA(\vecx)\bigr)\Bigr)\right]\cdot\vecsig
\end{equation}
are hermitian and traceless so that according to (\ref{spintranseq}) the 
spin transport matrices $\tilde d_\pm$ take values in $\SU (2)$. Since the 
solutions $(\vecP^\pm (t),\vecX^\pm (t))$ of Hamilton's equations of motion 
depend on the initial points $(\vecxi,\vecz^\pm)$, below we also write
$d_\pm (\vecxi,\vecz^\pm,t)$ for $\tilde d_\pm(\vecP^\pm(t),\vecX^\pm(t))$.
In these variables the composition law of the spin transport matrices 
reads
\begin{equation}
\label{spintranscomp}
d_\pm (\vecxi,\vecz^\pm,t+t')=d_\pm(\vecP^\pm(t'),\vecX^\pm(t'),t)\,
d_\pm (\vecxi,\vecz^\pm,t') \ .
\end{equation}

In order to calculate the leading semiclassical asymptotics of the time
evolution kernel $K(\vecx,\vecy,t)$ further one employs the method of
stationary phase to the integral (\ref{scFIO}). Due to the connection
of the phases $\phi^\pm$ with the solutions $S^\pm$ of the Hamilton-Jacobi
equations, the stationary points $\vecxi_{st}$ of $\phi^\pm$ are such that 
the canonical transformations generated by $S^\pm(\vecxi_{st},\vecx,t)$ map 
$(\vecp^\pm,\vecx)$ to $(\vecxi_{st},\vecy)$. These are therefore the end
and starting points, respectively, of the associated solutions 
$(\vecP^\pm(t'),\vecX^\pm(t'))$ of the equations of motion. Thus the
stationary points $\vecxi_{st}$ are in one-to-one correspondence with the
classical trajectories $\ga^\pm$ connecting $\vecy$ to $\vecx$ in time $t$. 
The expression that finally results from the application of the method of
stationary phase is to a large extent analogous to the Van Vleck-Gutzwiller
propagator (see \cite{Gut67}) for non-relativistic, spinless particles. 
Differences arise due to (i) the presence of two kinds of classical dynamics 
(generated by the two Hamiltonians $H^\pm$) instead of one, and (ii) the 
presence of the additional factors in the amplitude (\ref{a0sol}). The 
latter, however, need only be evaluated at the stationary points 
$\vecxi_{st}$. As a result one therefore obtains
\begin{equation}
\label{scprop}
K(\vecx,\vecy,t) = \frac{1}{(2\pi\ui\hbar)^{3/2}}\sum_{\ga^\pm}V_\pm(t)\,
d_\pm\,V_\pm^\dagger(0)\,D_{\ga^\pm}\,\ue^{\frac{\ui}{\hbar}R^\pm_{\ga^\pm}
-\ui\frac{\pi}{2}\nu_{\ga^\pm}}\,\bigl( 1+O(\hbar) \bigr) \ .
\end{equation}
Here the sum extends over all solutions $\ga^\pm$ of the equations of motion 
from $\vecy$ to $\vecx$ in time $t$, and $\nu_{\ga^\pm}$ denotes their
Morse indices. Moreover, the phase functions $\phi^\pm$ evaluated at the 
stationary point $\vecxi_{st}$ related to $\ga^\pm$ turn out to be
Hamilton's principal functions $R^\pm_{\ga^\pm}(\vecx,\vecy,t)=
S^\pm(\vecxi_{st},\vecx,t)-\vecxi_{st}\cdot\vecy$, and 
\begin{equation}
\label{Ddef}
D_{\ga^\pm} := \Bigl|\det\Bigl(-\frac{\partial^2 R^\pm_{\ga^\pm}}{\partial
x\partial y}(\vecx,\vecy,t)\Bigr)\Bigr|^{1/2}\ .
\end{equation}
We also introduced the abbreviation $V_\pm(t'):=V_\pm(\vecP^\pm(t'),
\vecX^\pm(t'))$.

Inspecting (\ref{scprop}) reveals a close similarity to the Van 
Vleck-Gutzwiller propagator. The summation goes over the classical 
trajectories of relativistic particles, such that here only the translational 
dynamics is relevant. The spin dynamics then enters through the additional 
factor $V_\pm(t)\,d_\pm\,V_\pm^\dagger(0)$ that describes the precession of 
a two-component spinor along the trajectory $\ga^\pm$ in the given external 
fields $\vecE$ and $\vecB$. In order to see this recall that 
$V_\pm^\dagger(\vecp,\vecx)$, being an isometry from a subspace in $\kz^4$ 
to $\kz^2$, projects four-component (Dirac) spinors to two-component spinors. 
These two components are the expansion coefficients of the projection to 
the eigenspace $P_0^\pm (\vecp,\vecx)\kz^4$ in the basis 
$\{e_1(\vecp,\vecx),e_2(\vecp,\vecx)\}$ or 
$\{f_1(\vecp,\vecx),f_2(\vecp,\vecx)\}$, respectively. The $\SU(2)$ matrix 
$d_\pm$ then propagates the two-component spinor along the trajectory 
$\ga^\pm$ towards its end point. The propagated two-component spinor is then 
converted back into a four-component (Dirac) spinor by $V_\pm(t)$. 
One can thus view the combined dynamics of the spin and translational
degrees of freedom as it enters the semiclassical propagator (\ref{scprop})
to be of a mixed quantum-classical type: the translational dynamics are
classical and drive the quantum mechanical spin dynamics. However, it 
can be shown that expectation values of the spin operator, which 
is proportional to $\vecsig$, in the two-component spinors exhibit the 
well known Thomas precession of a `classical' spin along $\ga^\pm$, see 
\cite{RubKel63,BolKep99,Spo00} for more details. Moreover, knowing the
classical spin precession, one can recover the full spin dynamics, as
described by the spin transport matrices $d_\pm$, with the help of
a certain dynamical and geometric phase \cite{BolKep99}.  
\section{The trace formula with expectation values}
\label{5sec}
Having the semiclassical propagator (\ref{scprop}) available, one could 
now proceed to derive a semiclassical trace formula, very much in
analogy to Gutzwiller's original trace formula for non-relativistic,
spinless particles \cite{Gut71,Gut90}. One only has to localise in energy
first, in order to project out the essential spectrum of the 
Dirac-Hamiltonian. To this end one chooses a smooth function $\chi\in
C_0^\infty(\rz)$ that is supported in the interval $(-mc^2,+mc^2)$ where 
the eigenvalues of $\hat H$ are located that one is interested in. Then 
$\chi(\hat H)$, defined by the functional calculus given by the spectral 
theorem, is a bounded and self-adjoint operator with pure point spectrum 
located in $\supp\chi$; its eigenvalues are $\chi(E_n)$. Hence the truncated 
time evolution operator $\hat U_\chi(t):=\ue^{-\frac{\ui}{\hbar}t\hat H}\,
\chi(\hat H)$ has a Schwartz kernel with spectral representation
\begin{equation}
\label{Kchispec}
K_\chi(\vecx,\vecy,t)=\sum_n\chi(E_n)\,\psi_n(\vecx)\psi_n^\dagger(\vecy)\,
\ue^{-\frac{\ui}{\hbar}E_n t}
\end{equation}
in terms of eigenspinors $\psi_n$ and eigenvalues $E_n\in\supp\chi$ of
$\hat H$. Since $K_\chi(\vecx,\vecy,t)=\chi(\hat H)\,K(\vecx,\vecy,t)$,
where $\chi(\hat H)$ acts on the variable $\vecx$, the leading semiclassical 
asymptotics of $K_\chi$ can immediately be concluded from (\ref{scprop}).
The only modification is provided by an additional factor of 
$\chi(E_{\ga^\pm})$, where $E_{\ga^\pm}$ denotes the energy of the trajectory 
$\ga^\pm$. The result can then be used to derive a semiclassical trace 
formula; see \cite{BolKep98,BolKep99}, where this procedure has been 
carried out.

Here, however, our aim is to include the expectation values $\langle\psi_n,
\hat B\psi_n\rangle$ of observables $\hat B$ as they have been introduced
in section \ref{3sec}. To this end one considers a suitably regularised 
trace of the operator $\hat B\,\hat U_\chi(t)$. The regularisation requires 
a test function $\rho\in C^\infty(\rz)$ with compactly supported Fourier 
transform, $\tilde\rho\in C_0^\infty(\rz)$, so that
\begin{equation}
\label{Uregdef}
\hat U_\chi [\tilde\rho] := \frac{1}{2\pi}\int_\rz\tilde\rho(t)\,\hat
U_\chi(t)\ \ud t
\end{equation}
defines a bounded operator. The map $\tilde\rho\mapsto\tr\hat U_\chi 
[\tilde\rho]$, where $\tr(\cdot)$ is the operator trace on $\cH$, then 
yields a tempered distribution, if the sum $\sum_n\chi(E_n)\rho(E_n/\hbar)$
is absolutely convergent. The latter condition certainly holds when the
eigenvalues do not accumulate in $\supp\chi$. One hence should choose 
$\chi$ in such a way that possible accumulation points of eigenvalues lie
outside of $\supp\chi$, and that $\chi$ vanishes sufficiently fast towards
accumulation points. Under this condition one can calculate the trace of the 
operator $\hat B\,\hat U_\chi[\tilde\rho\,\ue^{\frac{\ui}{\hbar}E(\cdot)}]$,
\begin{equation}
\label{pretrace} 
\tr\frac{1}{2\pi}\int_\rz\tilde\rho(t)\,\ue^{\frac{\ui}{\hbar}Et}\,\hat B\,
\hat U_\chi (t)\ \ud t = \sum_n\chi(E_n)\,\langle\psi_n,\hat B\psi_n\rangle
\,\rho\Bigl(\frac{E_n -E}{\hbar}\Bigr) \ .
\end{equation}
The right-hand side of (\ref{pretrace}) already provides the spectral side
of the trace formula we are aiming at. The semiclassical side of this
trace formula can then be obtained from the left-hand side upon introducing 
the kernel $K_\chi$ for $\hat U_\chi$ and calculating the trace in coordinate 
representation.

Before one can perform this computation, however, one needs the Schwartz
kernel for the operator $\hat B\,\hat U_\chi (t)$, or at least a
semiclassical expression for it. The latter can be derived from the
representation (\ref{scFIO}) of the time evolution kernel, when one employs 
the action of a Weyl operator $\hat D$ with symbol $D(\vecp,\vecx)$ on 
functions of the form $a(\vecx)\,\ue^{\frac{\ui}{\hbar}\phi(\vecx)}$. 
This reads
\begin{equation}
\label{WeylonWKB}
\bigl(\hat D\,a\,\ue^{\frac{\ui}{\hbar}\phi}\bigr) (\vecx) = \bigl(
D(\nabla\phi(\vecx),\vecx)\,a(\vecx) + O(\hbar) \bigr)\,
\ue^{\frac{\ui}{\hbar}\phi(\vecx)} \ ,
\end{equation}
see, e.g., \cite{Dui96}. In a first step one therefore has to apply the 
operator $\hat D=\chi(\hat H)$ to $K(\vecx,\vecy,t)$, which can be done 
with the help of (\ref{WeylonWKB}) since according to the functional 
calculus of \cite{Rob87} $\chi(\hat H)$ is a Weyl operator with principal 
symbol $D_0(\vecp,\vecx)=\chi(H(\vecp,\vecx))$. Then, in a second step, one 
repeats the procedure with the observable $\hat B$. Thus
\begin{equation}
\label{2Weylonamp}
\begin{split}
\hat B\,\chi(\hat H)\,a_0^\pm(\vecx,\vecy,t,\vecxi)\,
&\ue^{\frac{\ui}{\hbar}\phi^\pm(\vecx,\vecy,t,\vecxi)}\\
&=B_0\bigl(\nabla_x\phi^\pm(\vecx,\vecy,t,\vecxi),\vecx\bigr)\,\chi\bigl(
 H^\pm (\nabla_x\phi^\pm(\vecx,\vecy,t,\vecxi),\vecx)\bigr) \\
&\quad\ a_0^\pm(\vecx,\vecy,t,\vecxi)\,\ue^{\frac{\ui}{\hbar}\phi^\pm
 (\vecx,\vecy,t,\vecxi)} +O(\hbar) 
\end{split}
\end{equation}
yields the leading semiclassical asymptotics for the integrand of the 
Schwartz kernel for the operator $\hat B\,\hat U_\chi(t)$ in an analogous 
representation to (\ref{scFIO}). Calculating the trace on the left-hand 
side of (\ref{pretrace}) in coordinate representation therefore requires,
in leading semiclassical order, to evaluate the integrals
\begin{equation}
\label{traceint}
\begin{split}
\int_\rz\int_{\rz^3}\int_{\rz^3}
&\mtr\bigl(B_0(\nabla_x\phi^\pm,\vecx)\,a_0^\pm(\vecx,\vecx,t,\vecxi)\bigr)\\
&\tilde\rho(t)\,\chi\bigl(H^\pm(\nabla_x\phi^\pm,\vecx)\bigr)\,
 \ue^{\frac{\ui}{\hbar}(\phi^\pm(\vecx,\vecx,t,\vecxi)+Et)}\ 
 \ud\xi\,\ud x\,\ud t\ ,
\end{split}
\end{equation}
where $\mtr(\cdot)$ denotes a matrix trace.

Since the integration cannot be performed in closed form, one employs
the method of stationary phase. The connection of the phase with the 
generating functions $S^\pm$ then reveals that stationary points 
$(\vecxi_{st},\vecx_{st},t_{st})$ are such that $(\vecxi_{st},\vecx_{st})$ 
lies on a periodic orbit $\ga_p^\pm$ with energy $E$ and period $t_{st}$ 
of the classical dynamics generated by $H^\pm$. One therefore concludes
that up to terms of order $\hbar^\infty$ the expression (\ref{pretrace})
is given by a sum over the periodic orbits $\ga_p^\pm$ with periods
$T_{\ga_p^\pm}\in\supp\tilde\rho$. We recall that the method of stationary 
phase also demands that at stationary points the phase be non-degenerate. 
Here this means that transversal to the (connected) manifolds of stationary
points, which are at least one dimensional, the matrix of second derivatives 
of the phase with respect to $(\vecxi,\vecx,t)$ must have a (constant) rank 
of seven minus the dimension of the respective manifold. Let us explicitly
 mention two examples: 
\begin{enumerate}
\item
The entire hyper-surfaces
\begin{equation}
\label{Omegadef}
\Om_E^\pm := \{(\vecp,\vecx);\ H^\pm (\vecp,\vecx)=E\}
\end{equation}
of constant energy are manifolds of periodic points with trivial periods 
$t_{st}=0$. Thus $\Om^\pm_E\times\{t_{st}=0\}$
are five dimensional manifolds of stationary points. In this case the 
non-degeneracy condition means that $E$ is a regular value for the 
Hamiltonians $H^\pm$, i.e.\ $(\nabla_p H^\pm(\vecp,\vecx),\nabla_x 
H^\pm(\vecp,\vecx))\neq 0$ for all $(\vecp,\vecx)\in\Om_E^\pm$.  
\item
An isolated periodic orbit $\ga_p^\pm$ is non-degenerate, if its monodromy
matrix $\mz_{\ga_p^\pm}$, i.e.\ the restriction of the linearised
Poincar\'e map to the directions transversal to the orbit, has no eigenvalue
one.
\end{enumerate}
For simplicity, below we will always restrict attention to these cases, i.e.\ 
we require $E$ to be a regular value for the Hamiltonians and assume that
all periodic orbits with energy $E$ are isolated and non-degenerate. The 
latter condition is, e.g., fulfilled by hyperbolic periodic orbits, where 
all eigenvalues of $\mz_{\ga_p^\pm}$ are different from one in modulus. 
However, we would like to stress that a semiclassical trace formula can be 
derived under more general conditions. One only needs a non-degeneracy 
condition for the manifolds of stationary points in order to be able to 
apply the method of stationary phase.

For the explicit computation of the semiclassical side of the trace formula
one should treat each connected manifold of stationary points separately.
To this end we now introduce the partition of unity $\sum_j h_j(t) =1$,
with $h_j\in C_0^\infty(\rz)$,
under the integral (\ref{traceint}) such that $\supp h_j$ contains only
one period $T_{\ga_p^\pm}$ of a periodic orbit $\ga_p^\pm$ with energy
$E$; in particular, $h(T_{\ga_p^\pm})=1$. Then each summand, labeled by $j$,
corresponds to a contribution of exactly one (isolated and non-degenerate)
periodic orbit. We, furthermore, reserve the label $j=0$ for the contribution
of the manifolds $\Om_E^\pm$ for which $t_{st}=0$. The necessary 
computations are almost identical to the case of the trace formula without
expectation values of an observable and thus we can direct the reader
to \cite{BolKep99} for the details. 

It turns out that the leading semiclassical order of the Weyl term, i.e.\ 
of the contribution with $j=0$, reads
\begin{equation}
\label{Weylterm1}
\begin{split}
\chi(E)\,\frac{\tilde\rho(0)}{2\pi}\,\frac{1}{(2\pi\hbar)^2}
&\left[\int_{\rz^3}\int_{\rz^3}\mtr\bigl( B_0(\vecp,\vecx)\,P^+_0(\vecp,\vecx) 
 \bigr)\,\de\bigl(H^+(\vecp,\vecx)-E\bigr)\ \ud p\,\ud x \right.\\
&+\left.\int_{\rz^3}\int_{\rz^3}\mtr\bigl( B_0(\vecp,\vecx)\,
 P^-_0(\vecp,\vecx) 
 \bigr)\,\de\bigl(H^-(\vecp,\vecx)-E\bigr)\ \ud p\,\ud x \right] \ .
\end{split}
\end{equation}
Introducing the notation
\begin{equation}
\label{Liouville}
\ud\mu^\pm_E (\vecp,\vecx) := \frac{1}{\vol(\Om_E^\pm)}\,\de\bigl(
H^\pm(\vecp,\vecx)-E\bigr)\ \ud p\,\ud x
\end{equation}
for Liouville measure on $\Om_E^\pm$, the leading order of the Weyl term 
can also be given as
\begin{equation}
\label{Weylterm2}
\chi(E)\,\frac{\tilde\rho(0)}{2\pi}\,\frac{1}{(2\pi\hbar)^2}\,
\left[\vol(\Om_E^+)\int_{\Om_E^+}\mtr\bigl( B_0\,P_0^+\bigr)\ \ud\mu^+_E + 
\vol(\Om_E^-)\int_{\Om_E^-}\mtr \bigl( B_0\,P_0^-\bigr)\ \ud\mu^-_E \right] \ .
\end{equation}
Since the matrices $P_0^\pm$ are projectors, the integrands in 
(\ref{Weylterm2}) also read $\mtr (B_0\,P_0^\pm) = \mtr (P_0^\pm\,B_0\,
P_0^\pm) =:\mtr B_0^\pm$, and as a short-hand we will also use 
$\mtr\overline{B_0^\pm}^E$ for the integrals over $\Om_E^\pm$.

In the case of an isolated and non-degenerate periodic orbit $\ga_p^\pm$
the evaluation of its contribution to the semiclassical side of the
trace formula also proceeds in close analogy to the case without an
observable, until it comes to the point where an integration over the 
primitive periodic orbit associated with $\ga_p^\pm$ is required. In the 
present situation the object that has to be integrated along the primitive 
orbit is the matrix trace of 
\begin{equation}
\label{integrandpo}
B_0(\vecp,\vecx)\,V_\pm(\vecp,\vecx)\,d_\pm(\vecp,\vecx,T_{\ga_p^\pm})\,
V_\pm^\dagger(\vecp,\vecx) \ ,
\end{equation}
where the last three factors derive from the spin contribution to the
semiclassical propagator (\ref{scprop}). For the integration we parametrise
the periodic orbit by $(\vecP^\pm(t),\vecX^\pm(t))$, $0\leq t\leq 
T_{\ga_p^\pm}^\#$, where $T_{\ga_p^\pm}^\#$ denotes the primitive period
associated with $\ga_p^\pm$, and hence have to evaluate (\ref{integrandpo}) 
at $(\vecp,\vecx)=(\vecP^\pm(t),\vecX^\pm(t))$. The quantity that finally 
enters the trace formula then is a particular average of the principal symbol 
$B_0$ of the observable along the primitive periodic orbit, 
\begin{equation}
\label{B0pomean1}
\begin{split}
\mtr\overline{B_0}^{\ga_p^\pm} := \frac{1}{T_{\ga_p^\pm}^\#}
\int_0^{T_{\ga_p^\pm}^\#}\mtr\Bigl[
&B_0\bigl(\vecP^\pm(t),\vecX^\pm(t)\bigr)\,V_\pm\bigl(
 \vecP^\pm(t),\vecX^\pm(t)\bigr) \\ 
&d_\pm\bigl(\vecP^\pm(t),\vecX^\pm(t),T_{\ga_p^\pm}\bigr)\,
 V_\pm^\dagger\bigl(\vecP^\pm(t),\vecX^\pm(t)\bigr)\Bigr]\ \ud t \ .
\end{split}
\end{equation}
The integrand appearing in (\ref{B0pomean1}) can be interpreted as follows: 
the right-most term projects a four-component (Dirac) spinor at 
$(\vecP^\pm(t),\vecX^\pm(t))$ to the local $H^\pm$-eigenspace 
$P_0^\pm(\vecP^\pm(t),\vecX^\pm(t))\kz^4$ in a two-component representation. 
This two-spinor is then propagated once around the periodic orbit and 
subsequently converted back to a Dirac spinor which is finally mapped by 
$B_0$. The integration then averages the result of this procedure over the
primitive orbit related to $\ga_p^\pm$. 

Altogether, if all periodic orbits generated by the two Hamiltonians
$H^\pm$ with an energy $E$ that is not a critical value for these
Hamiltonians are isolated and non-degenerate, the semiclassical trace formula 
we have been aiming at reads
\begin{equation}
\label{STF}
\begin{split}
\sum_n\chi(E_n)\,
&\langle\psi_n,\hat B\psi_n\rangle\,\rho\Bigl(\frac{E_n -E}{\hbar}\Bigr) \\
&=\chi(E)\,\frac{\tilde\rho(0)}{2\pi}\,\frac{1}{(2\pi\hbar)^2}\,\Bigl(
 \vol(\Om_E^+)\,\mtr\overline{B_0^+}^E + \vol(\Om_E^-)\,\mtr\overline{B_0^-}^E
 \Bigr) + O(\hbar^{-1}) \\
&\quad+\sum_{\ga_p^\pm}\chi(E)\,\frac{\tilde\rho(T_{\ga_p^\pm})}{2\pi}\,
 \mtr\overline{B_0}^{\ga_p^\pm}\,A_{\ga_p^\pm}\,\ue^{\frac{\ui}{\hbar}
 S_{\ga_p^\pm}(E)}\,\bigl( 1+O(\hbar) \bigr) \ ,
\end{split}
\end{equation}
where as usual the action of a periodic orbit is denoted as
\begin{equation}
\label{action}
S_{\ga_p^\pm}(E) := \int_{\ga_p^\pm}\vecp\cdot\ud\vecx \ ,
\end{equation}
and the amplitude associated with a periodic orbit,
\begin{equation}
\label{amplitude}
A_{\ga_p^\pm} := \frac{T_{\ga_p^\pm}^\#\,\ue^{-\ui\frac{\pi}{2}
\mu_{\ga_p^\pm}}}{\bigl|\det\bigl(\mz_{\ga_p^\pm}-\eins_4\bigr)\bigr|^{1/2}}
\ ,
\end{equation}
incorporates the monodromy matrix $\mz_{\ga_p^\pm}$ and the Maslov index
$\mu_{\ga_p^\pm}$ of that orbit. 

Let us remark that this trace formula reduces to the one derived in
\cite{BolKep98,BolKep99} in the case $\hat B=\eins_\cH$. Obviously, on
the left-hand side of (\ref{STF}) the expectation value then disappears
and in the Weyl term $\mtr\overline{B_0^\pm}^E=2$. In order to see
what happens to the weights $\mtr\overline{B_0}^{\ga_p^\pm}$ of periodic
orbits we notice that after a cyclic permutation under the trace in 
(\ref{B0pomean1}) one can employ the fact that $V_\pm^\dagger V_\pm=\eins_2$,
and that $\mtr d_\pm(\vecp,\vecx,T_{\ga_p^\pm})=:\mtr d_{\ga_p^\pm}$
is independent of $(\vecp,\vecx)$; hence $\mtr\overline{B_0}^{\ga_p^\pm}=
\mtr d_{\ga_p^\pm}$. With these identifications the trace formula of 
\cite{BolKep98,BolKep99} is hence recovered.

As a final comment on the trace formula (\ref{STF}) let us mention that
one can view the left-hand side of (\ref{STF}) as resulting from an
application of the weighted spectral density 
\begin{equation}
\label{specdens} 
d_{\chi,B}(E) := \sum_n\chi(E_n)\,\langle\psi_n,\hat B\psi_n\rangle\,
\de (E-E_n)
\end{equation}
to the test function $\rho$ (after a suitable shift of the variable).
In this context the trace formula (\ref{STF}) can be converted into a 
distributional identity for the weighted spectral density (\ref{specdens}), 
\begin{equation}
\label{STFdense}
\begin{split}
d_{\chi,B}(E)
&=\chi(E)\,\frac{1}{(2\pi\hbar)^3}\,\Bigl(\vol(\Om_E^+)\,
 \mtr\overline{B_0^+}^E + \vol(\Om_E^-)\,\mtr\overline{B_0^-}^E \Bigr) 
 + O(\hbar^{-2}) \\
&\quad+\chi(E)\,\frac{1}{2\pi\hbar}\sum_{\ga_p^\pm}
 \mtr\overline{B_0}^{\ga_p^\pm}\,A_{\ga_p^\pm}\,\ue^{\frac{\ui}{\hbar}
 S_{\ga_p^\pm}(E)}\,\bigl( 1+O(\hbar) \bigr) \ .
\end{split}
\end{equation}
It is in this form that trace formulae are frequently presented (see,
e.g., Gutzwiller's original work \cite{Gut71}).
\section{Semiclassical averages of expectation values}
\label{6sec}
The left-hand side of the trace formula (\ref{STF}) can be seen as a weighted
superposition of the expectation values $\langle\psi_n,\hat B\psi_n\rangle$
corresponding to the eigenvalues $E_n$ in some interval about $E$ of
width proportional to $\hbar$. This is so since $\rho\in C^\infty(\rz)$
is a test function with Fourier transform $\tilde\rho\in C_0^\infty(\rz)
\subset\cS(\rz)$. Thus $\rho$ itself is, indeed, a Schwartz function,
i.e.\ it is rapidly decreasing, and therefore essentially cuts off all
terms corresponding to eigenvalues outside of some interval $[E-\hbar\om_1,
E+\hbar\om_2]$. Although the length of this interval shrinks to zero
in the semiclassical limit, the number of eigenvalues it contains grows,
since according to the Weyl term of the spectral density (compare 
(\ref{STFdense}) with $\hat B=\eins_\cH$) the mean density of eigenvalues
increases like $\hbar^{-3}$. In the following we therefore want to study 
a semiclassical average of the expectation values 
$\langle\psi_n,\hat B\psi_n\rangle$ related to the $N_I$ eigenvalues 
$E_n\in I(E;\hbar)=[E-\hbar\om,E+\hbar\om]$, see (\ref{Idef}). More 
precisely, we want to express the limit
\begin{equation}
\label{Szegoeaim}
\lim_{\hbar\to 0}\frac{1}{N_I}\sum_{E_n\in I(E;\hbar)}
\langle\psi_n,\hat B\psi_n\rangle 
\end{equation}
in terms of classical quantities.

It seems that the trace formula (\ref{STF}) is the ideal tool that allows
one to achieve the calculation of the limit (\ref{Szegoeaim}). For two
reasons, however, the situation is not so straight forward:
\begin{enumerate}
\item For the trace formula (\ref{STF}) we assumed that all periodic orbits 
be isolated and non-degenerate. This is a condition which is, e.g., met by 
hyperbolic classical dynamics. However, hyperbolicity is a strong chaotic
property which is not at all necessary for the result we are aiming at.
In fact, a considerably weaker condition suffices: it is only required
that the points on the energy shells $\Om_E^\pm$ that lie on periodic 
orbits are of Liouville measure zero. Any ergodic dynamical system shares
this property, but even many integrable ones as well.
\item For the test function $\rho$ in the trace formula one cannot simply 
choose the characteristic function $\chi_{[-\om,\om]}$ of the interval 
$[-\om,\om]$, which would provide the sharp cut-off (\ref{Szegoeaim}) in 
the summation over the eigenvalues, since $\chi_{[-\om,\om]}$ is not smooth 
and, moreover, its Fourier transform $\frac{2}{t}\,\sin\om t$ is not 
compactly supported. It even decreases too slowly to make the sum over 
periodic orbits in (\ref{STF}) convergent.
\end{enumerate}
Despite of these objections it is nevertheless possible to calculate the
limit (\ref{Szegoeaim}) in rather general situations. For the case of 
Pauli-Hamiltonians this has, indeed, been carried out in \cite{BolGla00}.
However, to achieve this one does not employ the trace formula (\ref{STF}) 
directly, but rather starts off from (\ref{pretrace}), which served as 
the basic relation from which the trace formula was derived. Again one 
has to calculate the integral (\ref{traceint}), but here we are only 
interested in the leading semiclassical order. By an inspection of the 
trace formula one realises that this is completely determined by the Weyl 
term, i.e.\ by the stationary points $(\vecxi_{st},\vecx_{st},t_{st})$ with
$t_{st}=0$ and $(\vecxi_{st},\vecx_{st})\in\Om_E^\pm$. As a non-degeneracy
condition we therefore only need that $E$ is not a critical value for
the two Hamiltonians $H^\pm(\vecp,\vecx)$. For later purposes we, however,
require this condition to hold for all $E'\in [E-\ve,E+\ve]$ with some 
$\ve>0$. It is then ensured that the non-trivial time components 
$t_{st}\neq 0$ of stationary points cannot accumulate at zero (see, e.g.,
\cite{Rob87}). We stress that no assumptions on the periodic orbits have
to be made apart from the condition that they form a measure-zero set on
the energy shells. Under the integral (\ref{traceint}) one now only needs a 
simple partition of unity, $1=h_0(t)+\bigl(1-h_0(t)\bigr)$, that separates
the stationary points $(\vecxi_{st},\vecx_{st},0)$ from those with 
$t_{st}\neq 0$. In order to estimate the latter contribution one exploits 
the fact that the points on periodic orbits, which are precisely the 
stationary points $(\vecxi_{st},\vecx_{st})$ with $t_{st}\neq 0$, are 
assumed to be of Liouville measure zero on $\Om_E^\pm$. As a consequence 
one obtains that the leading semiclassical order is solely determined by
the Weyl contribution (see, e.g., \cite{HelMarRob87,DimSjo99,BolGla00}), 
so that
\begin{equation}
\label{preSzegoe1}
\begin{split}
\sum_n\chi(E_n)\,
&\langle\psi_n,\hat B\psi_n\rangle\,\rho\Bigl(\frac{E_n -E}{\hbar}\Bigr) \\
&=\chi(E)\,\frac{\tilde\rho(0)}{2\pi}\,\frac{1}{(2\pi\hbar)^2}\,\Bigl(
 \vol(\Om_E^+)\,\mtr\overline{B_0^+}^E + \vol(\Om_E^-)\,\mtr\overline{B_0^-}^E
 \Bigr) + o(\hbar^{-2}) \ .
\end{split}
\end{equation}
In a next step one would like to get rid of the test function $\rho$
and to replace it by a sharp cut-off. Under the conditions mentioned above,
which led to the relation (\ref{preSzegoe1}), this can indeed be done with
the help of the Tauberian Lemma of \cite{BruPauUri95}. The only point that
still has to be clarified is that the Tauberian Lemma requires the
expectation values of $\hat B$ to be non-negative. Since, however, $\hat B$
is bounded, this condition can be satisfied after shifting the observable 
by a suitable constant. In effect, the Tauberian Lemma of \cite{BruPauUri95}
then allows to simply replace $\rho$ by the characteristic function 
$\chi_{[-\om,\om]}$, and its Fourier transform by $\frac{2}{t}\,\sin\om t$. 
We also recall that $E$ was chosen to be somewhere inside the gap 
$(-mc^2,+mc^2)$ of the essential spectrum of $\hat H$. Thus, for sufficiently 
small $\hbar$ the interval $I(E;\hbar)$ is completely contained in 
$(-mc^2,+mc^2)$ so that the function $\chi\in C^\infty_0(\rz)$ that serves 
as to cut off the essential spectrum can be chosen such that 
$\chi(E)=1=\chi(E_n)$ for all $E_n\in I(E;\hbar)$. Hence
\begin{equation}
\label{preSzegoe2}    
\sum_{E_n\in I(E;\hbar)}\langle\psi_n,\hat B\psi_n\rangle = 
\frac{\om}{\pi}\,\frac{1}{(2\pi\hbar)^2}\,\Bigl(\vol(\Om_E^+)\,
\mtr\overline{B_0^+}^E + \vol(\Om_E^-)\,\mtr\overline{B_0^-}^E\Bigr) + 
o(\hbar^{-2}) \ .
\end{equation}
Furthermore, the choice $\hat B=\eins_\cH$ in (\ref{preSzegoe2}) yields a
semiclassical representation for the number $N_I$ of eigenvalues in the
interval $I(E;\hbar)$,
\begin{equation}
\label{scN_I}    
N_I = \frac{\om}{\pi}\,\frac{2}{(2\pi\hbar)^2}\,\bigl(\vol(\Om_E^+)
+\vol(\Om_E^-)\bigr) + o(\hbar^{-2}) \ ,
\end{equation}
so that (\ref{preSzegoe2}) and (\ref{scN_I}) finally allow to determine 
the limit (\ref{Szegoeaim}), 
\begin{equation}
\label{Szegoe}
\lim_{\hbar\to 0}\frac{1}{N_I}\sum_{E_n\in I(E;\hbar)}
\langle\psi_n,\hat B\psi_n\rangle = \frac{1}{2}\,\frac{\vol(\Om_E^+)\,
\mtr\overline{B_0^+}^E +\vol(\Om_E^-)\,\mtr\overline{B_0^-}^E}
{\vol(\Om_E^+)+\vol(\Om_E^-)} \ .
\end{equation}
As desired, the right-hand side of this relation involves only classical
quantities.

The result (\ref{Szegoe}) can be converted into a statement about averages
of the Wigner transforms
\begin{equation}
\label{Wigner}
W[\psi_n](\vecp,\vecx) :=\int_{\rz^3}\ue^{-\frac{\ui}{\hbar}\vecp\cdot\vecy}
\,\overline{\psi_n}\bigl(\vecx-\tfrac{1}{2}\vecy\bigr)\otimes 
\psi_n\bigl(\vecx+\tfrac{1}{2}\vecy\bigr)\ \ud y
\end{equation}
of the eigenspinors $\psi_n\in I(E;\hbar)$, since expectation values of 
Weyl operators can be expressed as
\begin{equation}
\label{Wignerexpect}
\langle\psi_n,\hat B\psi_n\rangle = \frac{1}{(2\pi\hbar)^3}\int_{\rz^3}
\int_{\rz^3}\mtr\bigl(W[\psi_n](\vecp,\vecx)\,B(\vecp,\vecx;\hbar)\bigr)
\ \ud p\,\ud x\ .
\end{equation}
Here $\mtr (\cdot)$ denotes a matrix trace, whose presence prevents an
immediate conclusion from (\ref{Szegoe}). To proceed nevertheless, we remark 
that if one replaces the observable $\hat B$ by its diagonal part $\hat B_d$, 
see (\ref{Bsplit}) and (\ref{Bsplitdef}), the right-hand side of (\ref{Szegoe})
remains unchanged. This immediately follows from the form (\ref{Bsplitpsymb})
of the principal symbols of the diagonal part and the non-diagonal part,
respectively. Thus, in particular,
\begin{equation}
\label{Szegoend}
\lim_{\hbar\to 0}\frac{1}{N_I}\sum_{E_n\in I(E;\hbar)}
\langle\psi_n,\hat B_{nd}\psi_n\rangle = 0 \ . 
\end{equation}
We can therefore now restrict attention to observables $\hat B$ that are
Weyl quantisations of symbols $B=B_{0,d}=B_0^+ +B_0^-$. Starting with a
quantisation of, e.g., $B_0^+$ one splits this matrix valued symbol into 
a sum of contributions where all matrix entries but one vanish, so that 
now one essentially deals with scalar operators and symbols. Applying
(\ref{Szegoe}) to these then finally yields
\begin{equation}
\label{SzegoeWigner}
\begin{split}
\lim_{\hbar\to 0}
&\,\frac{1}{N_I}\sum_{E_n\in I(E;\hbar)}\frac{1}{(2\pi\hbar)^3}\,
 W[\psi_n](\vecp,\vecx) \\
&= \frac{1}{2}\,\frac{\vol(\Om_E^+)\,\de(H^+(\vecp,\vecx)-E)\,P_0^+ 
   (\vecp,\vecx)+\vol(\Om_E^-)\,\de(H^-(\vecp,\vecx)-E)\,P_0^-(\vecp,\vecx)}
   {\vol(\Om_E^+)+\vol(\Om_E^-)} \ .
\end{split}
\end{equation} 
Here the convergence as $\hbar\to 0$ has to be understood in the sense of
distributions, i.e.\ after an application to (matrix valued) test functions.
In (\ref{Szegoe}) the latter are given by the symbols of Weyl operators. 

Let us add a few remarks on the result (\ref{Szegoe}). Relations of this
kind, which express semiclassical averages of quantum observables in terms 
of classical averages are known as (weak versions of) {\it Szeg\"o limit
formulae} \cite{Gui79}. They are well known for Schr\"odinger-Hamiltonians
(see, e.g., \cite{HelMarRob87}) and have recently been established for 
Pauli-Hamiltonians \cite{BolGla00}. In the present case the classical side
contains the two contributions that arise from projections of the classical 
observable $B_0$ to the two eigenspaces of the symbol $H$ and hence
correspond to `particles' and `anti-particles', respectively. The relative
weights of the two terms are fixed by the relative volumes of the
corresponding energy shells. It often occurs that for a given energy $E$
only one of the energy shells is non-empty, and then only one contribution
appears on the right-hand side of (\ref{Szegoe}). The only exceptional
situations are those where the Klein paradox occurs, i.e.\ 
where a tunneling between particle and anti-particle states is possible.
The corresponding mixture is then accounted for by the presence of
both terms on the classical side of (\ref{Szegoe}). Due to the relation
(\ref{Szegoend}), however, semiclassically transitions from particle to 
anti-particle states, or vice versa, play no role. On the right-hand side 
of (\ref{SzegoeWigner}) this effect is caused by the fact that only terms
proportional to the projectors $P_0^\pm$ occur so that the semiclassically
averaged Wigner transforms are block-diagonal with respect to the
particle and anti-particle subspaces. 

Obviously, Szeg\"o limit formulae of the kind (\ref{Szegoe}) contain
considerably less information on expectation values than trace formulae
like (\ref{STF}). They demonstrate that the semiclassically leading term of 
the weighted spectral density (\ref{STFdense}) is the semiclassical average 
of the observable multiplied by the leading Weyl-term of the spectral 
density. Fluctuations of expectation values about the semiclassical
average are then described by the sum over periodic orbits in (\ref{STFdense})
and therefore depend on classical properties more sensibly than the average 
itself. However, the trace formula (\ref{STF}) does not allow to draw 
conclusions about individual expectation values since on its spectral side 
it contains the test function $\rho$, whereas on its semiclassical side the 
Fourier transform $\tilde\rho$ appears. Thus, by Fourier duality, shrinking 
the (effective) support of $\rho$ results in a growing (effective) support 
of $\tilde\rho$, finally leading to a divergent, and thus uncontrollable, sum 
over periodic orbits. The smallest support of $\rho$ that can, indeed, be 
handled leads to the semiclassical average (\ref{Szegoe}). However, in the 
same way as the Gutzwiller trace formula is not suited for the representation
of individual eigenvalues, but enables a semiclassical analysis of spectral 
correlations (see, e.g., \cite{Ber85,BolKep99a}), the trace formula 
(\ref{STF}) can be used to determine correlations of expectation values as, 
e.g., in \cite{EckFisKeaAgaMaiMue95}.

\vspace*{0.5cm}
\subsection*{Acknowledgment}
I would like to thank Rainer Glaser and Stefan Keppeler for useful 
discussions.

\end{document}